\input harvmac
\input epsf

\def\half{{1\over 2}}

\def\d{\Lambda}


\def\rdr{\rho^{rec}_\d}
\def\mr{\Omega_\d^{rec}}

\def\dr{\triangle \rho}
\def\tr{\tilde \rho}
\def\dk{\triangle k}
\def\rdm{\rho_\d^{min}}
\def\md{\Omega_\d^{min}}
\def\wdm{w_\d^{min}}

\def\br{\bar \rho}


\Title{}{\vbox{\centerline{Anthropic Principle Favors the Holographic
Dark Energy}}}

\centerline{Qing-Guo Huang$^1$ and Miao Li$^{1,2}$}

\medskip
\centerline{\it $^1$ Interdisciplinary Center of Theoretical
Studies} \centerline{\it Academia Sinica, Beijing 100080, China}
\medskip
\centerline{\it and}
\medskip
\centerline{\it $^2$ Institute of Theoretical Physics}
\centerline{\it Academia Sinica, P. O. Box 2735} \centerline{\it
Beijing 100080}
\medskip

\centerline{\tt huangqg@itp.ac.cn} \centerline{\tt mli@itp.ac.cn}
\medskip

We discuss the anthropic principle when applied to the holographic
dark energy. We find that if the amplitude of the density
fluctution is variable, the holographic dark energy fares better
than the cosmological constant. More generally, the anthropic
predictions agree better with observation for dark energy with
$w_\d=p_\d/\rho_\d$ decreasing over time.

\Date{Oct. 2004}

\nref\ex{A. G. Riess et al., Astron. J. 116 (1998) 1009,
astro-ph/9805201; S. Perlmutter et al., Astrophys. J. 517 (1999)
565, astro-ph/9812133; A. G. Riess et al., Astrophys. J. 607
(2004) 665, astro-ph/0403292; Y. Wang and M. Tegmark, Phys. Rev.
Lett 92 (2004) 241302, astro-ph/0403292. }
\nref\wmap{C. L.
Bennett et al., Astrophys. J. Suppl. 148 (2003) 1,
astro-ph/0302207. }

\nref\sw{S. Weinberg, Phys. Rev. Lett. 59 (1987) 2607.}

\nref\msw{H. Martel, P. R. Shapiro, S. Weinberg, Astrophys. J. 492
(1998) 29, astro-ph/9701099. }

\nref\ys{Y. S. Piao, hep-th/0407258.}

\nref\vc{A. Aguirre, Phys. Rev. D. 64 (2001) 083508,
astro-ph/0106143; M. J. Rees, astro-ph/0401424; M. Tegmark and M.
J. Rees, Astrophs. J. 499 (1998) 526, astro-ph/9709058. }

\nref\ghjw{M. L. Graesser, S. D. H. Hsu, A. Jenkins and M. B.
Wise, hep-th/0407174.}

\nref\mt{M. Tegmark, astro-ph/0410281. }

\nref\ml{Miao Li, hep-th/0403127. }

\nref\bek{J. D. Bekenstein, Phys. Rev. D 7, 2333 (1973); J. D.
Bekenstein, Phys. Rev. D 9, 3292 (1974); J. D. Bekenstein, Phys.
Rev. D 23, 287 (1981); J. D. Bekenstein, Phys. Rev. D 49, 1912
(1994); S. W. Hawking, Phys, Rev. D 13, 191 (1976); G. 't Hooft,
gr-qc/9310026; L. Susskind, hep-th/9409089, J. Math. Phys 36, 6377
(1994). }

\nref\ckn{A. Cohen, D. Kaplan and A. Nelson, hep-th/9803132, Phys.
Rev.  Lett. 82 (1999) 4971; S. Thomas, Phys. Rev. Lett.  89 (2002)
081301. }

\nref\hsu{S. D. H. Hsu, hep-th/0403052.}

\nref\holo{Y. Gong, hep-th/0404030; M. Ito, hep-th/0405281;  S.
Hsu and A. Zee, hep-th/0406142; E. V. Shuryak, hep-th/0407190; K.
Ke and M. Li, hep-th/0407056; R. Horvat, astro-ph/0404204; B. Wang
and E. Abdalla and R. K. Su, hep-th/0404057.}

\nref\hl{Qing-Guo Huang, Miao Li, JCAP 0408 (2004) 013,
astro-ph/0404229. }

\nref\hg{Qing-Guo Huang, Yungui Gong, JCAP 0408 (2004) 006,
astro-ph/0403590. }

\nref\sl{K. Enqvist and M. S. Sloth, hep-th/0406019; K. Enqvist,
S. Hannestad and M. S. Sloth, astro-ph/0409275; J. Shen, B. Wang,
E. Abdalla and R. Su, hep-th/0412227; Y. Myung, hep-th/0412224; Y.
Gong, B. Wang and Y. Zhang, hep-th/0412218. }

\nref\peebles{P. J. E. Peebles, Astrophys. J. 147 (1967) 859.}

The discovery of dark energy \refs{\ex, \wmap} can be regarded as
one of the most significant discoveries in observational cosmology
in the past several years. Although recent observations seem to
indicate that the cosmological constant is a good candidate for
dark energy, there is still plenty of room for imaginations, and
the nature of dark energy remains one of the deepest mysteries
crying for understanding and explanation. The problem of dark
energy includes: Why the order of magnitude of the dark energy
density is much smaller than quantum corrections in quantum field
theory? And why it is the same order as today's critical energy
density of our universe? The latter is called the cosmic
coincidence problem.

Weinberg in 1987 proposed that the anthropic principle can be a
good resolution to the cosmological constant problem \sw, and some
people even take his calculation as a prediction for the existence
of a non-vanishing dark energy. The anthropic principle presumes
that the problem of dark energy can never be posed if there are no
astronomers in a subuniverse at all. Thus, according to this idea,
the density of dark energy can take a range of values in different
subuniverses and the probability distribution of its observed
values is conditioned by the requirement that there must be
someone to measure it. As a very interesting result, the anthropic
argument predicts the existence of a small positive cosmological
constant \refs{\sw, \msw}, whose value is not far from the
observed one.

In the inflation scenario, our observed universe is a tiny patch
in the early universe and the whole space-time may contain many
disconnected regions which will grow up to many subuniverses. This
offers a framework for applying the anthropic principle. Recent
developments in the string/M theory compactifications also favor
the working of the anthropic principle, after all, there are too
many vacua for us to select a unique or a few based on physical
principles. It appears that the anthropic rule is the only hope to
select a vacuum in the vast string theory landscape. On the other
hand, there could be  some dynamical processes connecting
different vacua, for example see \ys, again, the anthropic
principle seems unavoidable.

However, the authors of \vc\ found that life may exist in
universes with some of the cosmological parameters differing from
those in our own subuniverse by several orders of magnitude, if
larger density fluctuations than those in our universe are
allowed. The quantitative analysis of this proposal is carried out
in \ghjw, and the result obtained suggests that the anthropic
prediction for the cosmological constant is much weakened by
allowing the variation of the amplitude of the primordial
fluctuation power spectrum, a much larger cosmological constant is
more favorable. The anthropic over-prediction of the cosmological
constant is also investigated in \mt. If the cosmological constant
is really a constant, and all the assumptions involved in the
calculation of \refs{\ghjw, \mt} hold, this is really bad news for
people believing that the anthropic principle is the resolution.

The purpose of the
present note is to examine the same question in the framework of the
holographic dark energy \ml. Since the holographic dark energy evolves in time,
it is not surprising that our result differs significantly, and we find that the
anthropic principle makes a better prediction here if dark energy is holographic.
Put in another way, the
anthropic argument favors the holographic dark energy, not the cosmological
constant.

The non-extensive entropy bound $S \leq \pi M^2_p L^2$ on the
total entropy $S$ in a volume $L^3$ proposed by Bekenstein et al.
\bek\ suggests that quantum field theory breaks down in large
scale. A. Cohen et al. \ckn\ proposed a relationship between UV
and IR cut-offs to rescue the local quantum field theory from the
spelling of formation of black holes, this results in a up-bound
on the zero-point energy density. We may view this as the
holographic dark energy, the magnitude of this dark energy is
consistent with the cosmological observations. However, Hsu
recently pointed out that the equation of state is not correct for
describing the accelerating expansion of our universe \hsu. Very
recently, one of us \ml\ suggested that we should use the proper
size of the future event horizon of our universe to cut-off the
large scale, resulting in a correct equation of state of dark
energy. The density of the holographic dark energy in a spatially
flat universe is introduced in \ml\ (For previous work, see \ckn;
for related work, see \refs{\holo-\hg}; and for a work relating
the holographic dark energy to low multipoles in CMB spectrum, see
\sl) \eqn\dfhd{\rho_\d=3 d^2 M_p^2 R_h^{-2},} where $M_p$ is the
reduced Planck mass, $R_h$ is the proper size of the future event
horizon \foot{Causality in the holographic dark energy model was
discussed in \ml\ and cannot be understood completely. }
$R_h=a(t)\int^{\infty}_t dt' /a(t')$ and $d$ is a free parameter.
In particular, the holographic dark energy behaves as a
cosmological constant with $w_\d=-1$ as $t \rightarrow \infty$ for
d=1. The equation of state of the holographic dark energy
\refs{\ml,\hl} is \eqn\shd{w_\d=-{1\over 3} {d \ln \rho_\d \over d
\ln a}-1 =-{1\over 3}\left(1+{2\over d}\sqrt{\Omega_\d} \right), }
and the evolution of $\Omega_\d$ in a universe with matter and
dark energy is governed by (as resulting from the Friedmann
equation) \eqn\dg{{d \Omega_\d \over d \ln a} = \Omega_\d
(1-\Omega_\d) \left(1+{2\over d}\sqrt{\Omega_\d}\right).} Solving
this equation \hg, we obtain \eqn\ma{\eqalign{&\ln \Omega_\d - {d
\over 2+d}\ln (1-\sqrt{\Omega_\d})+{d \over 2-d} \ln
(1+\sqrt{\Omega_\d})-{8 \over 4-d^2} \ln (d+2\sqrt{\Omega_\d})\cr
&=-\ln (1+z)+y_0. }}

Weinberg obtained an inequality between the density fluctuation
and the cosmological constant in \sw. Given an over-density
$\delta$, the cosmological constant can not be greater than a value
depending on $\delta$ in order for this region to collapse to form a galaxy or
a cluster of galaxies. In our case, the density of the holographic
dark energy evolves during the evolution of universe, the result of \sw\
can not be used, we need to
find the bound for the holographic dark energy independently.
Following Peebles \peebles\ and Weinberg \sw, the density
fluctuation in a universe with holographic dark energy can be
modelled as a sphere within which there is a uniform excess density
$\dr(t)$, and a gravitational field described by FRW metric with
positive curvature constant $\dk>0$ and the scale factor $a(t)$.
The evolution of the fluctuation is governed by the Friedmann equation
\eqn\evpt{{\dot a}^2+\dk={1\over 3M_p^2}a^2(\rho+\dr+\rho_\d),}
where $\rho$ is the unperturbed cosmic mass density. The total matter
density observes the mass conservation law thus
\eqn\ecr{a^3 (\rho+\dr)=const.}
In general as $t\rightarrow 0$, we have
$\rho_\d \ll \dr \ll \rho$. To the first order in the perturbation
solution of the Friedmann equation, there is a relation between the perturbed curvature constant
and density perturbation, the same as in \sw
\eqn\pcc{\dk={5\over 9 M_p^2}a^2 (\rho+\dr)^{2/3} \tr^{1/3}, }
where \eqn\dtr{\tr={(\dr)^3 \over \rho^2},}
a quantity independent of time.

The R.H.S. of eq. \evpt\ reaches its minimum when
\eqn\mcd{{\rho+\dr \over
\rdm}={1-\md\over \md}=-(3\wdm+1).} Using eq. \shd, we get
\eqn\om{\left(1+{2\over d}\sqrt{\md} \right) \md=1,}
where quantities with superscript min are the ones at the time
when the R.H.S. of \evpt\ reaches its minimum.
In this paper we take a special case with $d=1$, then $\md \simeq
0.432$ and $\wdm \simeq -0.772$.

Now, for this region to re-collapse, there must be a time when
$\dot{a} =0$. For this to happen, the minimum of the R.H.S. of eq.
\evpt\ should be less than $\dk$. Thus we find a bound for the
holographic dark energy \eqn\bhd{{1\over 3M_p^2} {-3\wdm \over
(3\wdm+1)^{2/3}}a^2 {(\rdm)}^{1/3}(\rho+\dr)^{2/3}<\dk.} Using eq.
\pcc, the bound can be expressed as \eqn\bnhd{\rdm < \left({5\over
3} {(3\wdm+1)^{2/3} \over -3\wdm} \right)^3 \tr \simeq 0.645 \tr.}
For the cosmological constant case, $w_\d=-1$ and $\rho_\d < {500
\over 729} \tr$, the same as \sw.

We like to find a bound on the holographic energy density at the
recombination time. The holographic dark energy density is not
constant. Using eq. \shd\ and \dg, the relation between the
holographic dark energy density at the recombination time $\rdr$
and $\rdm$ is found to be \eqn\recrm{\eqalign{\ln {\rdr \over
\rdm}=-2\left({1-d \over 2+d} \ln {1-\sqrt{\mr} \over
1-\sqrt{\md}}+{1+d \over 2-d} \ln {1+\sqrt{\mr} \over
1+\sqrt{\md}} \right. \cr \left.-{12 \over 4-d^2} \ln
{d+2\sqrt{\mr} \over d+2\sqrt{\md}}+\ln {\mr \over \md} \right).}}
Since $\mr \simeq \rdr/\br \ll 1$, where $\br$ is the energy
density of non-relativistic matter at recombination, for $d=1$, we
have \eqn\rerm{\rdm = \left(1+\sqrt{\md}\right)^{-4}
\left(1+2\sqrt{\md}\right)^{8} (\md)^{-2} \left({\rdr \over \br}
\right)^3 \br \simeq 585 \left({\rdr \over \br} \right)^3 \br. }
Combining eq.\bnhd\ and \rerm, the bound of the energy density of
the holographic dark energy at recombination is found to be
\eqn\bnhdc{{\rdr \over \br}< 0.103 \delta, } where $\delta =
{\triangle \rho \over \br}|_{rec}$ is the value of the average
fractional over-density at recombination. However, the bound for
the cosmological constant is ${\rho_\d \over \br} \leq {500 \over
729} \delta^3$. The anthropic bound for holographic dark energy is
quite different from the cosmological constant case.

We assume that the geometry in a sub-universe in general is
different from others and there should be a distribution of the
holographic dark energy in the ensemble. According to the
principle of Bayesian statistics, the probability distribution of
the dark energy is
\eqn\probd{P_{obs}(\rho_\d)={A(\rho_\d)P(\rho_\d) \over
\int^{\infty}_0 A(\rho_\d)P(\rho_\d) d \rho_\d },} where
$A(\rho_\d)$ is the mean number of astronomers who can measure the
dark energy density in a subuniverse. Since $\rho_\d$ is much
smaller than energy scales in theory of elementary particle
physics, we usually assume that the {\it a priori} probability
distribution $P(\rho_\d)$ is a constant (This point can be
questioned if dark energy is not really a constant, since the
underlying dynamics of the nonconstant dark energy may result in a
nontrivial distribution for $\rho_\d$) . The authors of \msw\
propose that the number of astronomers $A(\rho_\d)$ who can
measure $\rho_\d$ in a subuniverse should be proportional to the
fraction $F(\rho_\d)$ of baryonic matter incorporated in galaxies.
We also adopt these assumptions, thus, the fraction of mass
winding up in galaxies for the holographic dark model is \eqn\fm{F
(x=\rdr/\br)=\int^{\infty}_{x \over 0.103} d \delta {\delta
N(\delta) \over {x \over 0.103} \times {1 \over s}+ \delta}, }
where we assume the fluctuation distribution be Gaussian
distribution \eqn\dn{N(\delta)={1 \over \sigma} \left({2 \over
\pi} \right)^{1/2} \exp \left(-{\delta^2 \over 2 \sigma^2}
\right), } and $s$ is the ratio of the volume of the over-density
sphere to the volume of the under-density shell surrounding the
sphere. In this paper we set $s=1$. The relation \fm\ is the same
as in the one in \msw, except that the lower limit of the integral
is set by the new inequality \bnhdc. This lower limit is quite
different from the old one, since ours is a linear relation
between $\delta$ and $x= \rho_\d/\bar{\rho}$, while the old one is
a linear relation between $\delta$ and $x^3$. This difference is
the main source of our improved result over that of \ghjw.

To compute the probabilities, we need to normalize \fm.
The normalization integral in eq. \fm\ can be calculated by
interchanging the orders of integrations over $x$ and $\delta$:
\eqn\norm{\int^{\infty}_{0} F(x) d x = 0.103 \times \ln
2 \left({2 \over \pi} \right)^{1/2} \sigma. }
Thus for a given
energy density of the holographic dark energy $\rdr=x \br$ at
recombination, the normalized probability distribution becomes
\eqn\pobs{P_{obs}(x)= {1 \over 0.103 \times \ln 2 \left({2 \over
\pi} \right)^{1/2} \sigma} \int^{\infty}_{x \over 0.103} d \delta
{\delta N(\delta) \over {x \over 0.103} + \delta}. }
And the
probability of a holographic dark energy with energy density greater
than $\rho_{\d *}^{rec}$ is \eqn\pobsg{P_{obs}(x>x_*)={1
\over \ln 2 \times \sigma^2} \int^{\infty}_{x_* \over 0.103}
\delta e^{-{\delta^2 \over 2 \sigma^2}} \ln \left({0.206 \delta
\over x_* + 0.103\delta} \right) d \delta, } where $x_* = \rho_{\d
*}^{rec}/\br$. Using $\Omega_{\d *}^0=0.75$ for $d=1$ in our
subuniverse, we fix the parameter $y_0$ to be $-1.67$ in eq.
\ma. As a consequence, we find that at recombination with $z_{rec}^* = 1089$
(the first year result of WMAP \wmap)  $x_* \simeq
\Omega_{\d *}^{rec} \simeq 1.821 \times 10^{-4}$ in our
subuniverse.

In order to calculate the probability distribution, we need to
know the variance $\sigma$ as in $N(\delta)$. We follow \msw,
\eqn\fs{\sigma^2={1 \over 2 \pi^2}
\int^{\infty}_{0}P(k)W^2(kR)k^2dk, } where $P(k)$ is the power
spectrum at recombination, $W(x)=\exp (-x^2/2)$ is a Gaussian
window function and $R$ is the minimal size of structures that are
relevant in providing environment for emergence of intelligence.
The window function is used to filter the underlying density field
to exclude the contribution from perturbations of small
wavelengths. We do not really know the sufficient conditions for
the formation of astronomers, thus we assume the order of the
magnitude of $R$ be around 1 Mpc. After taking the evolution of
the power spectrum into account, the authors of \msw\ find
\eqn\szr{\sigma(z_{rec}) = (c_{100})^{(n+3)/2} (1+z_{rec})^{-1}
\Gamma^{(n+3)/2} \delta_H f(\Omega_\d^0) K_n^{1/2} (q_{max}), }
where $c_{100}=2997.9$ is the speed of light in units of 100
km/sec, $n$ is the index of the primordial power spectrum (here we
assume a scale-invariant spectrum with $n=1$), \eqn\dh{\delta_H =
1.94 \times 10^{-5}({\Omega_{m}^0})^{-0.785-0.05\times \ln
\Omega_{m}^0}} is COBE normalized amplitude of fluctuations at
horizon crossing, and \eqn\tk{\eqalign{&f(\Omega_\d^0) = {6
(\Omega_\d^0)^{5/6} \over 5 (\Omega_m^0)^{1/3}}
\left[\int^{\Omega_\d^0 / \Omega_m^0}_0 {d w \over w^{1/6}
(1+w)^{3/2}} \right]^{-1}, \cr &\Gamma=\Omega_m^0 h
e^{-\Omega_B^0-\Omega_B^0/\Omega_m^0}, \cr &q={k \over h \Gamma
\hbox{Mpc} ^{-1}}, \cr &K_n(q_{max})=\int^{\infty}_0 q^{n+2}
T^2(q) W^2 \left(2\pi {q \over q_{max}}\right) d q, \cr &T(q)={\ln
(1+2.34q)\over2.34q}
[1+3.89q+(16.1q)^2+(5.46q)^3+(6.71q)^4]^{-1/4}, } } where
$q_{max}={2\pi / (h \Gamma R \hbox{Mpc})}$.

We take $\Omega_{B *}^0=0.044$, $h_*=0.71$,
$\Omega_{m*}^0=1-\Omega_{\d *}^0=0.25$ in our subuniverse \wmap\
\hg. As in \msw, we first assume that the variance is the same in
different subuniverses (this was latter challenged in \vc\ and
\ghjw). Numerically $\sigma_*=2.397\times 10^{-3}$ for $R= 1$ Mpc,
with this input, we find $P_{obs}(x>x_*)=28.2 \%$; for $R=2$ Mpc,
$\sigma_*=1.736\times 10^{-3}$, $P_{obs}(x>x_*)=16.4 \%$. Both
results are satisfying according to the ``principle of
mediocrity".

Graesser et al. recently proposed, following \vc,  that there is
{\it a priori} probability distribution for $\sigma$ \ghjw. They
worked with an inflation model with potential of the inflaton
$V(\phi)=\lambda \phi^p$ ($p \geq 2$), assuming the distribution
of the coupling constant $\lambda$ be flat. The primordial power
spectrum is proportional to the coupling constant $\lambda$. The
power spectrum at recombination $P(k)$ in eq. \fs\ should be
proportional to $\lambda$ also, therefore we have $\sigma^2
\propto \lambda$. For a slow-roll inflation, $\lambda$ is a small
parameter, we may assume a constant {\it a priori} probability
distribution of $\lambda$. \foot{The same argument could just as
well be applied to $\lambda^2$ and the results will be different.
} The probability distribution of $\sigma$ should be
\eqn\pds{P(\sigma) \propto {d \lambda \over d \sigma} \sim
\sigma.} The probability distribution for $\rho_\d = x \br$
becomes \eqn\hwpr{P_{obs}(x) \sim \int d\sigma P(\sigma)
F(x,\sigma).} After normalization, we have
\eqn\hwp{P_{obs}(x>x_*)={1 \over \ln 2 \times
\int^{\sigma_{max}}_{\sigma_{min}} \sigma d\sigma}
\int^{\sigma_{max}}_{\sigma_{min}} {d\sigma \over \sigma}
\int^{\infty}_{x_* \over 0.103} \delta e^{-{\delta^2 \over 2
\sigma^2}} \ln \left({0.206 \delta \over x_* + 0.103\delta}
\right) d \delta.} In the holographic dark energy model, we find
$P_{obs}(x>x_*)=79.7 \%$ if $\sigma \in [0.1 \sigma_*,10
\sigma_*]$, $P_{obs}(x>x_*)=97.5 \%$ if $\sigma \in [0.01
\sigma_*,100 \sigma_*]$, for $R=1$ Mpc; and $P_{obs}(x>x_*)=73.6
\%$ if $\sigma \in [0.1 \sigma_*,10 \sigma_*]$,
$P_{obs}(x>x_*)=96.6 \%$ if $\sigma \in [0.01 \sigma_*,100
\sigma_*]$, for $R=2$ Mpc. However for the cosmological constant,
we have $P_{obs}(x>x_*)=99.96 \%$ for $\sigma \in [0.1 \sigma_*,10
\sigma_*]$ when $R=1$ Mpc; and $P_{obs}(x>x_*)=99.90 \%$ for
$\sigma \in [0.1 \sigma_*,10 \sigma_*]$ when $R=2$ Mpc. The
``principle of mediocrity" presumes that our universe is a typical
one among those friendly to intelligence, thus probability
$P_{obs}(x>x_*)$ must be a number not too close to 1 or too close
to 0. Comparing our result about the holographic dark energy to
the one on the cosmological constant, it appears that the
anthropic rule favors the holographic dark energy after we take a
$priori$ probability distribution for $\sigma$ into account.

The reason that the result on the holographic dark energy looks better
than that on the cosmological constant is that the holographic dark energy
evolves rather rapidly in the matter dominated era, so around the
recombination time, dark energy is greater, and the lower bound
on the mass perturbation to form structures is much larger-because
of both the larger dark energy density and the modified lower bound relation.

The last point we want to discuss about is whether the {\it a
priori} distribution of dark energy is flat or not. For a
cosmological constant, we can imagine of no reason for a
nontrivial distribution near zero. However, the holographic dark
energy evolves, it may have a nontrivial distribution. According
to eq. \shd, $w_\d \simeq -1/3$ if a component of energy other
than the dark energy dominates the universe and $\rho_\d \propto
a^{-2}$. We assume that there was an inflation period and the
inflation is driven by a inflaton with a potential $V(\phi)=\half
m^2 \phi^2$. The COBE normalization results in $m=7.4\times
10^{-6}$, here we always work with the unit in which the reduced
Planck mass $M_p=1$. During the inflation era, the holographic
dark energy may start with a value much greater than inflaton
energy. The slow-roll inflation begins when the inflaton energy is
comparable to the dark energy, and the latter is quickly
red-shifted away. The holographic dark energy density at the end
of inflation is roughly \eqn\hdinf{\rho^{end}_\d \sim \half m^2
\phi_i^2 e^{-2N_{tot}}, } where $\phi_i$ is the initial value of
$\phi$ and $N_{tot} \simeq \phi_i^2/4-1/2$ is the total e-folding
number. We see that the holographic energy really depends on the
number of e-folds, and it will have a nontrivial distribution if
$N_{tot}$ is a random number. It is quite reasonable that we
require the initial energy density of inflaton be less than
$M_p^4$, which means $\phi_i^2 \leq 2/m^2 \simeq 3.7 \times
10^{10}$. On the other hand, the condition for the end of
inflation is $\phi^2 \simeq 2$. Thus we have $2 \leq \phi_i^2 \leq
3.7 \times 10^{10}$, and $\ln \rho_\d^{end} \sim -\phi_i^2/2$.
Since $\rdr \propto \rho_\d^{end}$, \eqn\prr{d\phi_i \sim {d \rdr
\over \rdr \sqrt{|\ln \rdr |} }. } Assuming the {\it a priori}
probability distribution of $\phi_i$ be a constant, we find
probability distribution of dark energy \eqn\propdp{P(\rho_\d)\sim
{1 \over \rdr \sqrt{|\ln \rdr |}}.} This probability distribution
favors small $\rdr$, the integral over $\rdr$ diverges near the
origin, so there must be a cut-off for $\rdr$. This cut-off
corresponds to the upper bound on the initial value $\phi_i$ which
in turn derives from the upper bound on the initial inflaton
energy, thus, we have $(\rho_\d^{end})_{min} \simeq 10^{-8\times
10^9}$, an extremely small value. Multiplying the weight function
in eq. \propdp, we find that the anthropic argument favors an
extremely small value of dark energy and this prior is ruled out
be the observed larger $\rho_\d$ at very high significance.

However, it is rather premature to discuss the nontrivial distribution of
dark energy, since we do not know whether the flat measure $d\phi_i$ is correct,
and since unknown physics in the inflation era and beyond
can not be ignored.

To summarize, we find that if the amplitude of density fluctuation
is variable, the anthropic consideration favors the holographic
dark energy over the cosmological constant. It is possible that
the holographic dark energy is not the only form of dark energy
that fares better than the cosmological constant. A more general
conclusion is that, the anthrophic predictions agree better with
observation for a the state parameter $w_\d$ decreasing over time.

Acknowledgments.

The research of Huang is supported by a grant from NSFC and a
grant from China Postdoctoral Science Foundation. Li is supported
by a grant from CAS and a grant from NSFC.

\listrefs
\end